\documentclass[12pt]{iopart}

\usepackage{graphicx,color}
\usepackage{caption}
\usepackage{subcaption}
\usepackage{wrapfig}
\usepackage{iopams}

\begin{document}

\title{The interaction of intense ultrashort laser pulses with cryogenic He jets}

\author{M. Shihab $^{1,2}$, T. Bornath $^1$, and R. Redmer $^1$}

\address{ $^1$ Institut f\"ur Physik, Universit\"at Rostock, D-18051 Rostock, Germany}
\address{ $^2$ Department of Physics, Faculty of Science, Tanta University, Tanta 31527, Egypt}
\ead{mohammed.shihab@uni-rostock.de, mohammed.shihab@science.tanta.edu.eg}
\vspace{10pt}
\begin{indented}
\item[]\today
\end{indented}

\begin{abstract}

We study the interaction of intense ultrashort laser pulses with cryogenic He jets using 2d/3v relativistic Particle-in-Cell simulations (XOOPIC).
Of particular interest are laser intensities $(10^{15}-10^{20})$~W/cm$^2$, pulse lengths $\le 100$~fs, and the frequency regime $\sim 800$~nm 
for which the jets are initially transparent and subsequently not homogeneously ionized. 
Pulses $\ge 10^{16}$~W/cm$^2$ are found to drive ionization along the jet and outside the laser spot, 
the ionization-front propagates along the jet at a fraction of the speed of light.
Within the ionized region, there is a highly transient field, which may be interpreted as 
two-surface wave decay and as a result of the charge-neutralizing disturbance at the jet-vacuum interface.
The ionized region has solid-like densities and temperatures of few to hundreds of eV, i.e., 
warm and hot dense matter regimes.
Such extreme conditions are relevant for high-energy densities as found, e.g., 
in shock-wave experiments and inertial confinement fusion studies.
The temporal evolution of the ionization is studied considering theoretically a pump-probe x-ray Thomson scattering (XRTS) scheme.
We observe plasmon and non-collective modes that are generated in the jet, and their amplitude is proportional to the ionized volume. 
Our theoretical findings could be tested at free-electron laser facilities such as FLASH and the European XFEL (Hamburg) 
and the LCLS (Stanford).
\end{abstract}

%
\vspace{2pc}

\section{Introduction}

Material under extreme conditions --known as, warm (WDM) and hot (HDM) dense matter-- is located between 
conventional condensed matter states and ideal plasmas. 
Neither condensed matter models nor plasma models are suited to describe this intermediate region correctly. 
Moreover, analytical models such as perturbation techniques become increasingly complicated due to partial 
ionization, electron degeneracy, strong correlations, bound-state level shift, and pressure ionization~\cite{Kraeft,Vinko2012,Preston2013}.
The rigorous determination of the equation of state, the radiation opacity, and the transport properties
of WDM and HDM is mandatory for controlling the complex sequence of inertial confinement fusion (ICF)
experiments~\cite{Remington2006,Moses,Committee}, designing specific compression paths in shock-wave experiments, 
and developing interior and evolution models for, e.g., giant planets~\cite{Knudson2012,Swift2012,French2012}. 

Warm and hot dense matter can be generated and investigated in the laboratory via intense ultrashort laser 
pulses provided by free electron laser (FEL) facilities such as FLASH and European XFEL in Hamburg and LCLS in Stanford.
The state-of-the-art is to perform pump-probe experiments, where a pump pulse heats the target up to tens of 
electronvolts and compresses it up to solid-state density and beyond. 
To investigate the dynamic processes on such short time scales, the warm or hot dense matter is probed by a 
second pulse at well-defined time delays with respect to the pump pulse. The probe pulse operates at frequencies 
higher than the plasma frequency and thereby penetrates the target and scatters off electrons.
Therefore, x-ray Thomson scattering (XRTS)~\cite{Glenzer2009} is a robust and flexible experimental diagnostic 
tool for high-power laser-plasma interactions~\cite{Ross2010, Glenzer1999}, WDM 
studies~\cite{Lee2009,Kritcher2009,Kritcher2011,Fortmann2012,White2013,Souza2014}, 
astrophysics~\cite{Taylor1994,Visco2012,Regan2012,Remington,Fortney}, 
ICF plasmas and shock-wave experiments~\cite{Froula2007,Gregori2008,Kritcher2008}.

Different parameters control the energy deposited in the target and, consequently, determine the ionized volume, 
the electron temperature, the ion temperature, and the degree of ionization. 
The wavelength of the pump pulse and the geometry of the target are examples of such parameters.
The interaction of optical lasers with H and He droplets has been studied theoretically using particle-in-cell 
techniques~\cite{Theile2010,Sperling2013,Liseykina2013}.
Compressed and inhomogeneous overdense plasmas are produced. Ionization inside the interior of the droplet
was observed due to the penetration of the laser into the droplet. This could be interpreted
in terms of an enhanced electric field at the droplet surface --as predicted by the Mie theory-- 
and a plasma wave propagating from there into the target.

In this paper, we focus on another set up: The interaction of optical laser pulses with He jets. 
Flexible cryogenic liquid jets have been used in XRTS experiments~\cite{Ulf2014,Glenzer2016}. 
Understanding the ionization dynamics and kinetic instabilities during and after the pump pulse 
is necessary to interpret the XRTS spectrum. In general, the interaction of intense lasers with 
metals and overdense plasmas has widely been investigated for potential 
applications~\cite{Bulanov,Macchi2002,Macchi2001,Quinn2009,Tokita2011,Nakajima2013,Gopal2013,Fedeli2016}, 
e.g., plasmonics~\cite{Maier} and high harmonic generation~\cite{Teubner}.
The interaction of intense laser pulses with a metallic wire was investigated theoretically and 
experimentally~\cite{Tokita2015}. A strong transient field propagating at the speed of light along 
the wire was observed, nevertheless, only a small number of fast electrons is propagating with velocities comparable to the speed of light.
The transient field is confined at the surface and evanescent in the normal to the interface in a distance proportional to the skin depth.
Transverse magnetic (TM) modes might be coupled to the surface electrons producing surface waves over long 
distances away from the interaction area.

In this work, we use a different set up replacing the metal wire with a neutral cryogenic He jet. 
In this case of a dielectric jet, an inhomogeneous, possibly overdense, plasma is formed in the interaction area.
Each overdense layer oscillates with the driving field and radiates a transient field with harmonics related to its density.
When the field has an energy greater than the ionization threshold, ionization takes place at the surface and within the interior of the jet, i.e., the field is not well trapped at the surface.
We performed simulations for the interaction of s- and p-polarized laser light with He jets.
Both polarizations lead qualitatively to the same results. Pulses with intensities $\ge 10^{16}$~W/cm$^2$ 
generate overdense plasmas and drive ionization along the jet and outside the laser spot.
The ionization front propagates at a fraction of the speed of light. 

The paper is organized as follows. Section 2 is devoted to the findings of the numerical simulations for the interaction of intense laser radiation with He jets.
Especially, it explains the plasma generation in He jets of $0.5$~$\mu$m and $5$~$\mu$m diameter.
Section 3 discusses possibilities to verify the theoretical findings experimentally by x-ray Thomson scattering. We present synthetic spectra for the collective as well the non-collective scattering regimes.
At the end we give conclusions.

\section{Plasma generation in He jets}

The interaction of optical laser pulses with liquid He jets is investigated utilizing the Xoopic 
code which is a 2d3v particle-in-cell code. 
The relativistic equation of motion is solved with Maxwell's equations self-consistently. 
The laser energy is delivered to the target via tunneling ionization.
Unless otherwise stated, the impact ionization is switched-off. 
Test runs showed that for the intensities considered here the effect of collisional 
ionization during the laser pulse is much smaller than that of field ionization \cite{Liseykina2013,Sperling2015}. 

\subsection{Pumping jets with a diameter of 0.5~$\mu$m}\label{A}

As a first case, we consider a jet diameter of 0.5~$\mu$m, s-polarized laser pulses with a pulse length of 
$20$~fs (the envelop is a half-width for Gaussian), a Gaussian shape with a full width at half maximum (FWHM) 
of $1.5$~$\mu$m, and a wavelength of $800$~nm which irradiate He jets with a semi-infinite length normally. 
The pulse comes from the left and propagates along the x-axis in the vacuum until it reaches the jet.
The jet is placed parallel to the y-axis with initial density $2.2\times 10^{22}$~cm$^{-3}$ and temperature of 20~K.
Different simulations for laser intensities from $10^{15}$~W/cm$^2$ to $10^{19}$~W/cm$^2$ have been carried out.
To minimize the numerical instabilities and to ensure the reproducibility of the results, we assumed different 
grid sizes from $\Delta x= \Delta y = \lambda/200$ to $\Delta x= \Delta y = \lambda/50$, time steps from 
$\Delta t=0.07 \Delta x/c$ to $\Delta t=0.6 \Delta x/c$  where $c$ is the speed of light, and the number of 
super-particles per cell from 60 to 120. All the configurations produce qualitatively the same results. 
Therefore, we show unless otherwise stated the results belonging to $\Delta x= \Delta y = \lambda/200$, 
$\Delta t=0.07 \Delta x/c$, and 120 super-particles per cell.

\begin{figure}[!tbp]
  \centering
  \begin{minipage}[b]{0.45\textwidth}
   \includegraphics[width=\textwidth]{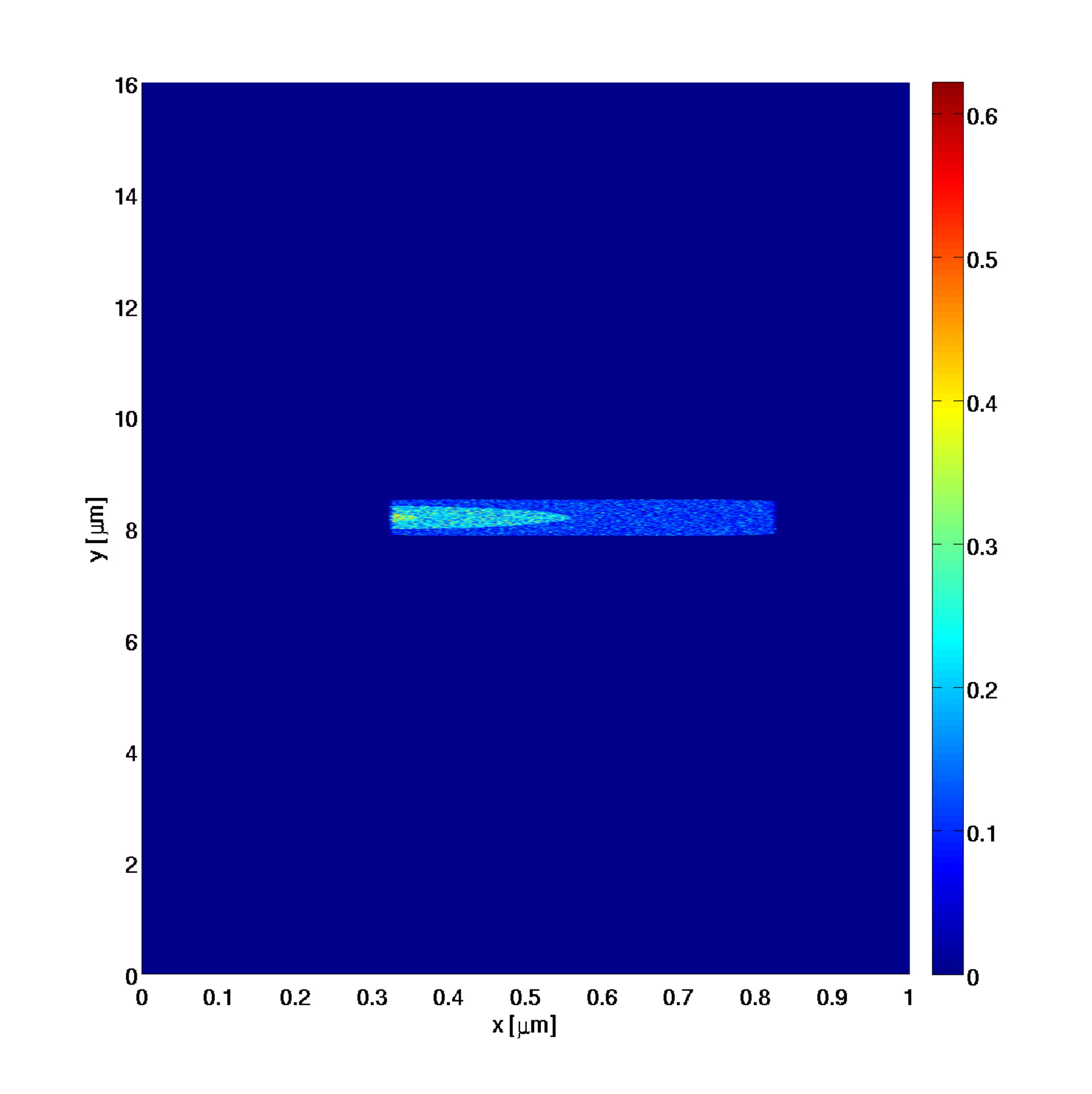}
    \subcaption{}
      \end{minipage}
  \hfill
  \begin{minipage}[b]{0.45\textwidth}
  \includegraphics[width=\textwidth]{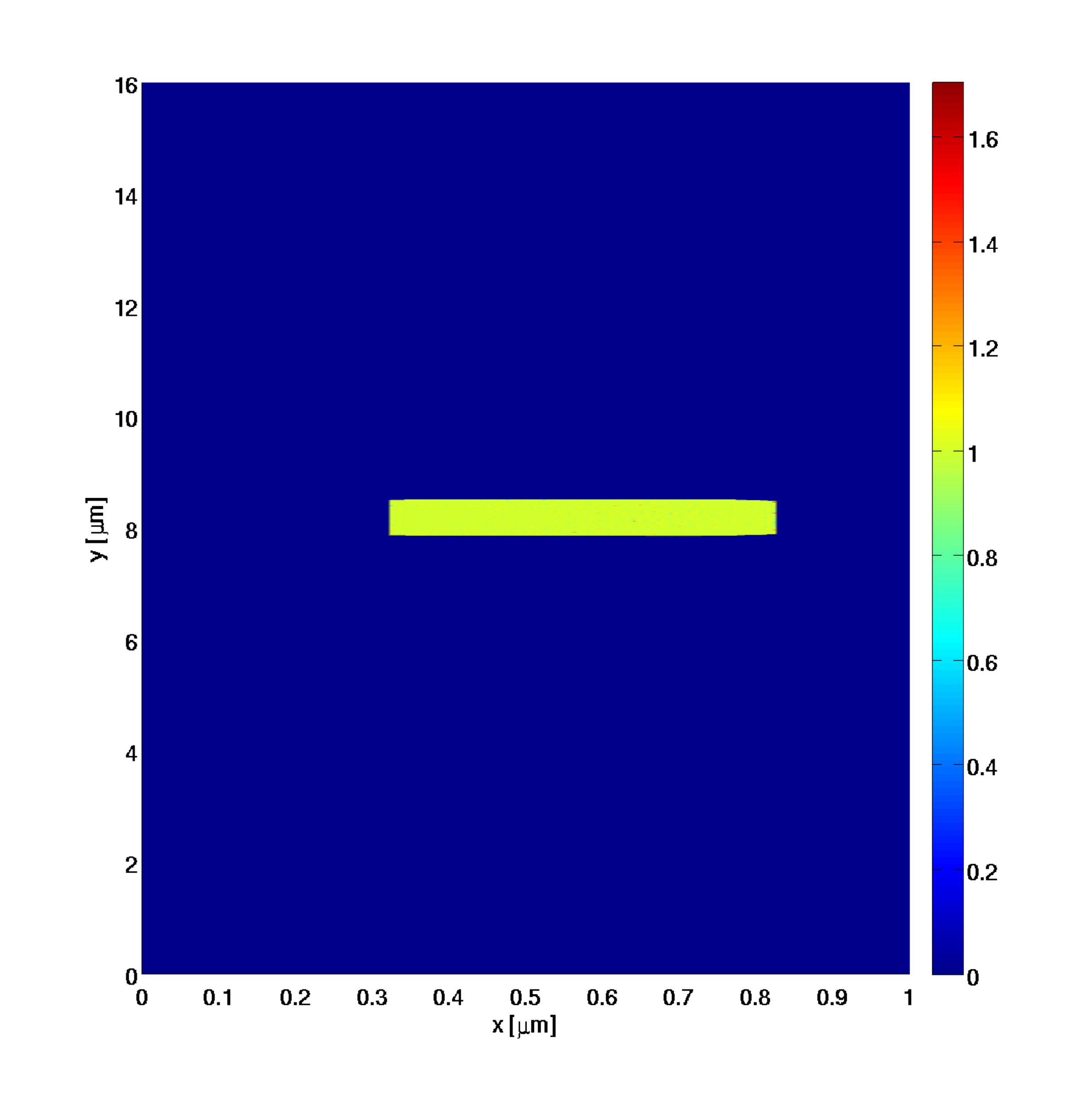}
    \subcaption{}
  \end{minipage}
  \caption{The electron density in units of critical density (a) and the degree of ionization (b) at 20~fs. The laser intensity is $10^{15}$~W/cm$^2$.}
  \label{neZ15}
\end{figure}

First, we show the electron density and the degree of ionization at the end of the laser pulse, i.e., at $20$~fs
for the intensity $10^{15}$~W/cm$^2$, see \figurename{~(\ref{neZ15}~a)} and \figurename{~(\ref{neZ15}~b)}, respectively.
Only a small area inside the laser spot (i.e., interaction area) is ionized. An underdense plasma with a degree of 
ionization close to $1$ is formed. As clearly can be seen from \figurename{~(\ref{neZ15}~b)}, the laser energy is not 
sufficient to produce He$^{+2}$ ions and the deposited energy is restricted to a small area around the pulse 
peak.

%

%

\begin{figure}[!tbp]
  \centering
  \begin{minipage}[b]{0.45\textwidth}
  \includegraphics[width=\textwidth]{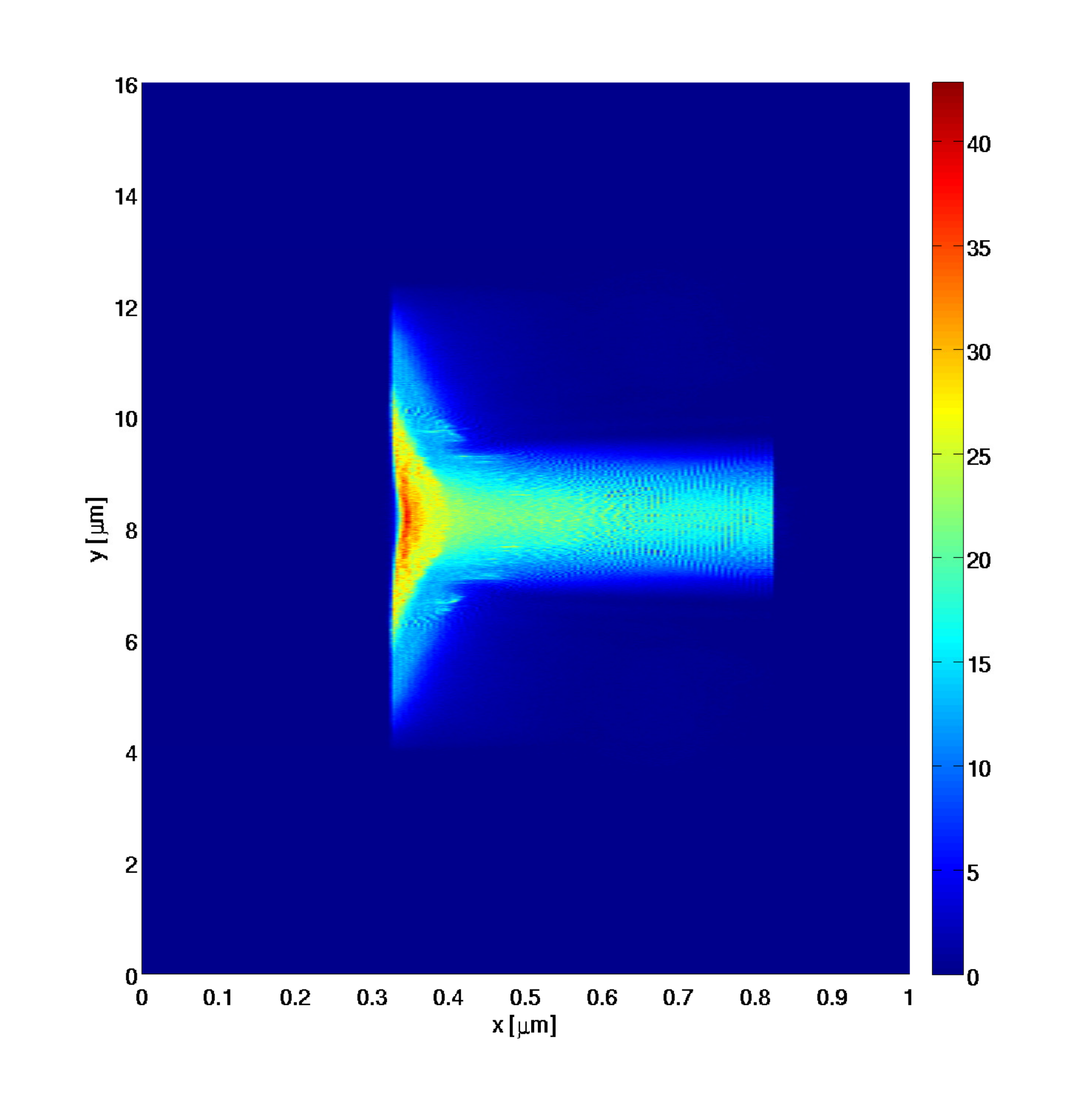}
   \subcaption{}
      \end{minipage}
  \hfill
  \begin{minipage}[b]{0.45\textwidth}
  \includegraphics[width=\textwidth]{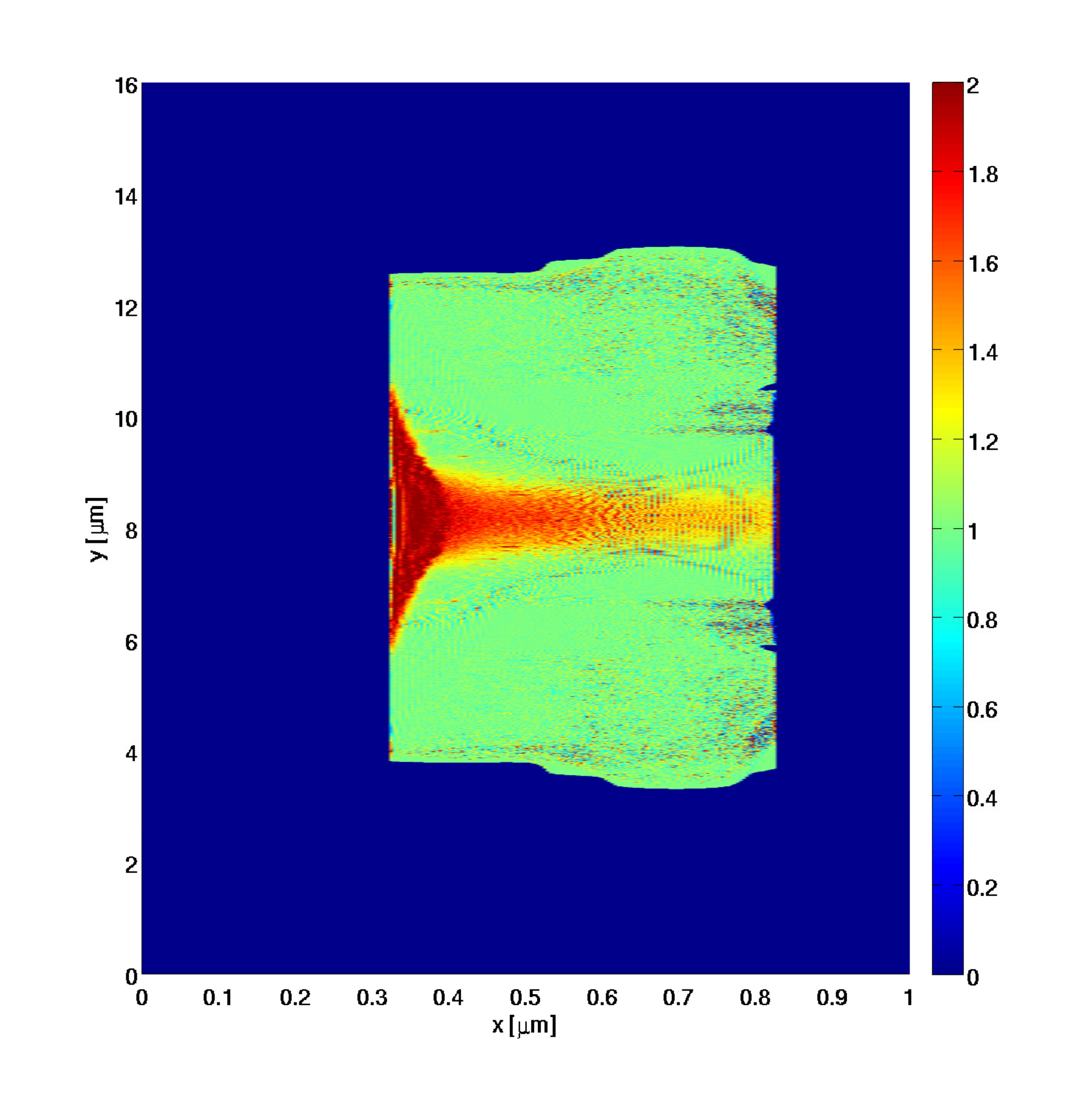}
   \subcaption{}
  \end{minipage}
  \caption{The electron density in units of critical density (a) and the degree of ionization (b) at 20~fs. The laser intensity is $10^{18}$~W/cm$^2$.}
  \label{neZ18}
\end{figure}


\begin{figure}
\center
\resizebox{0.5\textwidth}{!}{\includegraphics{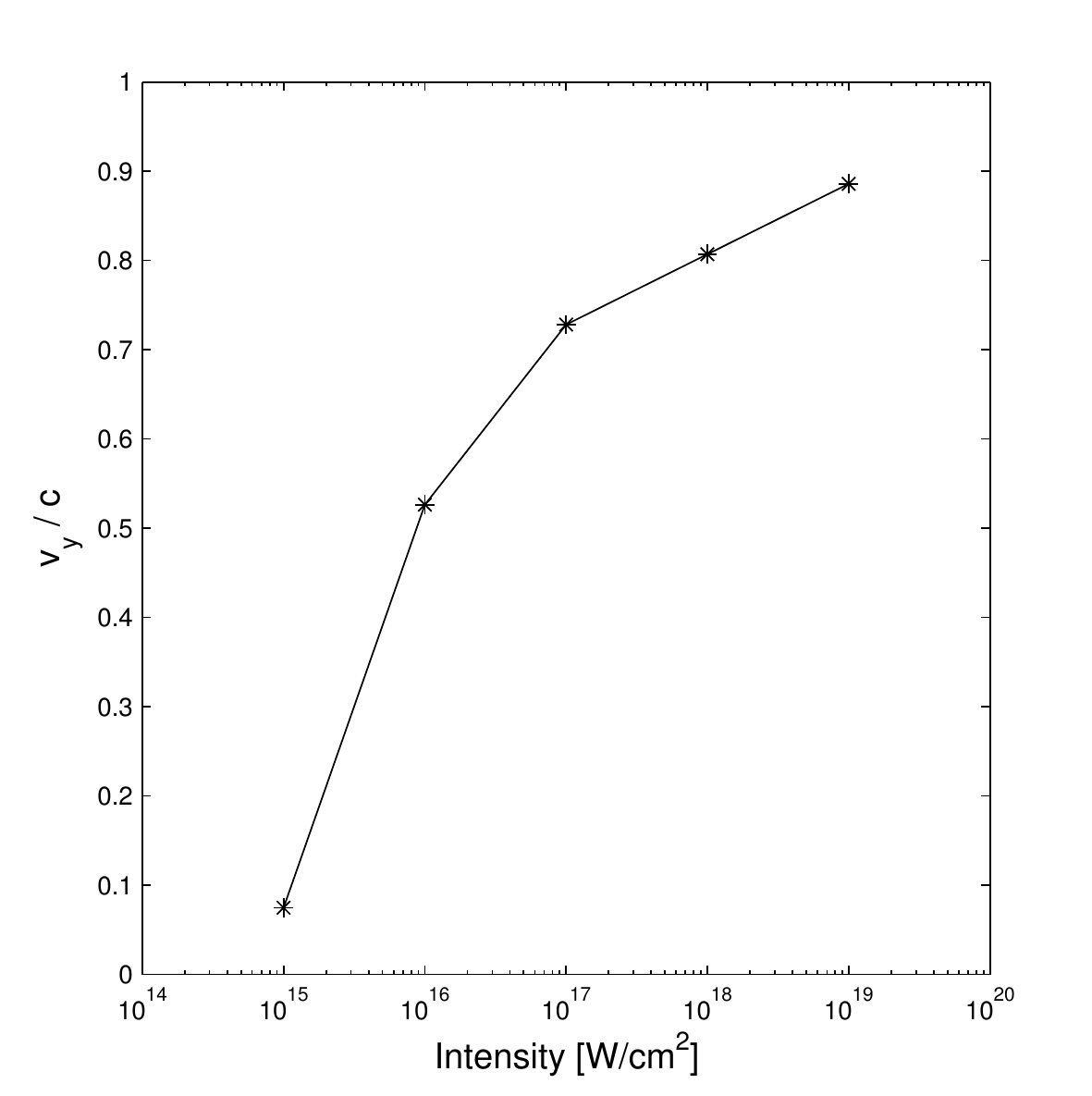}}
\caption{The average speed of the ionization front at the end of the laser pulse (20~fs) along the jet in units of the speed of light versus the laser intensity.}
\label{Vya}
\end{figure}

On the other hand, the electron density and the degree of ionization corresponding to the
laser intensity of $10^{18}$~W/cm$^2$ are shown in \figurename{~(\ref{neZ18}~a)} and \figurename{~(\ref{neZ18}~b)}, respectively. 
An overdense plasma is formed; the electron density at the surface reaches 40 times the critical density. 
The generated plasma is spatially inhomogeneous. The electron density at the near side is greater than the electron 
density at the rear side. Therefore, \figurename{~(\ref{neZ18}~a)} shows implicitly the localization of the energy at 
the near side of the jet.
In \figurename{~(\ref{neZ18}~a)}, one can easily distinguish between two areas with different degrees of ionization. 
The laser spot area which is ionized by direct interaction with the laser pulse exhibits a degree of ionization close 
to $2$, i.e., the concentration of He$^{+2}$ is dominant. Outside the laser spot area, plasma with a lower degree of 
ionization is formed. 
In addition, it is worth reporting that the Fourier analysis of the field inside the ionized area provides the 
existence of localized high harmonics with respect to the driving laser frequency. Moreover, the six components 
of the electromagnetic field exist and all components exhibit a wave-like structure.

The laser pulse consists of a number of subcycles. 
When the intensity of the first few subcycles can produce overdense plasmas, the next subcycles in the laser pulse 
will excite a counterpropagating transient electric field along the jet in a process similar to the two-plasmon 
decay process discussed by Macchi \textit{et al.}~\cite{Macchi2002,Macchi2001}.
The transient field will ionize the jet outside the laser spot.
Hence, the ionized area at the end of the laser pulse depends on the time at which the overdense plasma is generated.
The average speed of the ionization front along the jet at the end of the laser pulse is shown in \figurename{~(\ref{Vya})}.
In this context, the average speed is half of the length of the ionized area along the jet at the end of the laser 
pulse over the pulse length. There is an offset in the ionized area at the initial time due to the Gaussian of the laser pulse.
Therefore, the propagation speed of the ionization front would be smaller than the average speed shown in  \figurename{~(\ref{Vya})}.
The average speed is a fraction of the speed of light. 
We want to emphasize the difference between the interaction of high intense laser pulses ($\ge 10^{18}$~W/cm$^2$) 
with dielectric and neutral jets and metals or pre-ionized overdense plasmas.  In metals and homogeneously pre-ionized 
overdense plasmas with step-density profile, the laser pulse excites a strong transient field at the interface (the surface) instantaneously. 
This field is confined at the surface and evanescent in the normal to the interface in a distance proportional to the 
skin depth. Transverse magnetic (TM) modes might be coupled to the surface electrons producing surface waves over long 
distances away from the interaction  area~\cite{Bulanov,Macchi2002,Macchi2001,Quinn2009,Tokita2011,Nakajima2013,Gopal2013,Fedeli2016}.
The field oscillations may decay parametrically into two counterpropagating electron surface waves (labeled + and -, respectively), where the three-wave process holds:
\begin{equation}\label{k_SW}
 k_{\rm l}=k_+ + k_- ,
\end{equation}
\begin{equation}\label{w_SW}
 \omega_{\rm l}=\omega_+ + \omega_-.
\end{equation}
Here, $\omega_{\rm l}$ and $k_{\rm l}$ are the frequency and the wavenumber of the driving field, respectively. $\omega_{\pm}$ and $k_{\pm}$ are the frequencies and the wavenumbers of the surface waves.
Note, preimposed target modulation or grating is not required to satisfy eqs. (\ref{k_SW}) and (\ref{w_SW}).
On the other hand, in dielectric jets an inhomogeneous overdense plasma is formed in the interaction area, hence, the criterion of the two surface plasmon decay is matched.
Each overdense layer oscillates with the driving field and radiates a transient field with harmonics related to its density.
When the field has an energy greater than the ionization threshold, ionization takes place. Moreover, the field can propagate 
also to the interior of the jet, i.e., it is not well trapped at the surface.

\subsection{Pumping jets with a diameter of 5~$\mu$m}\label{B}

\begin{figure}[!tbp]
  \centering
  \begin{minipage}[b]{0.45\textwidth}
    \includegraphics[width=\textwidth]{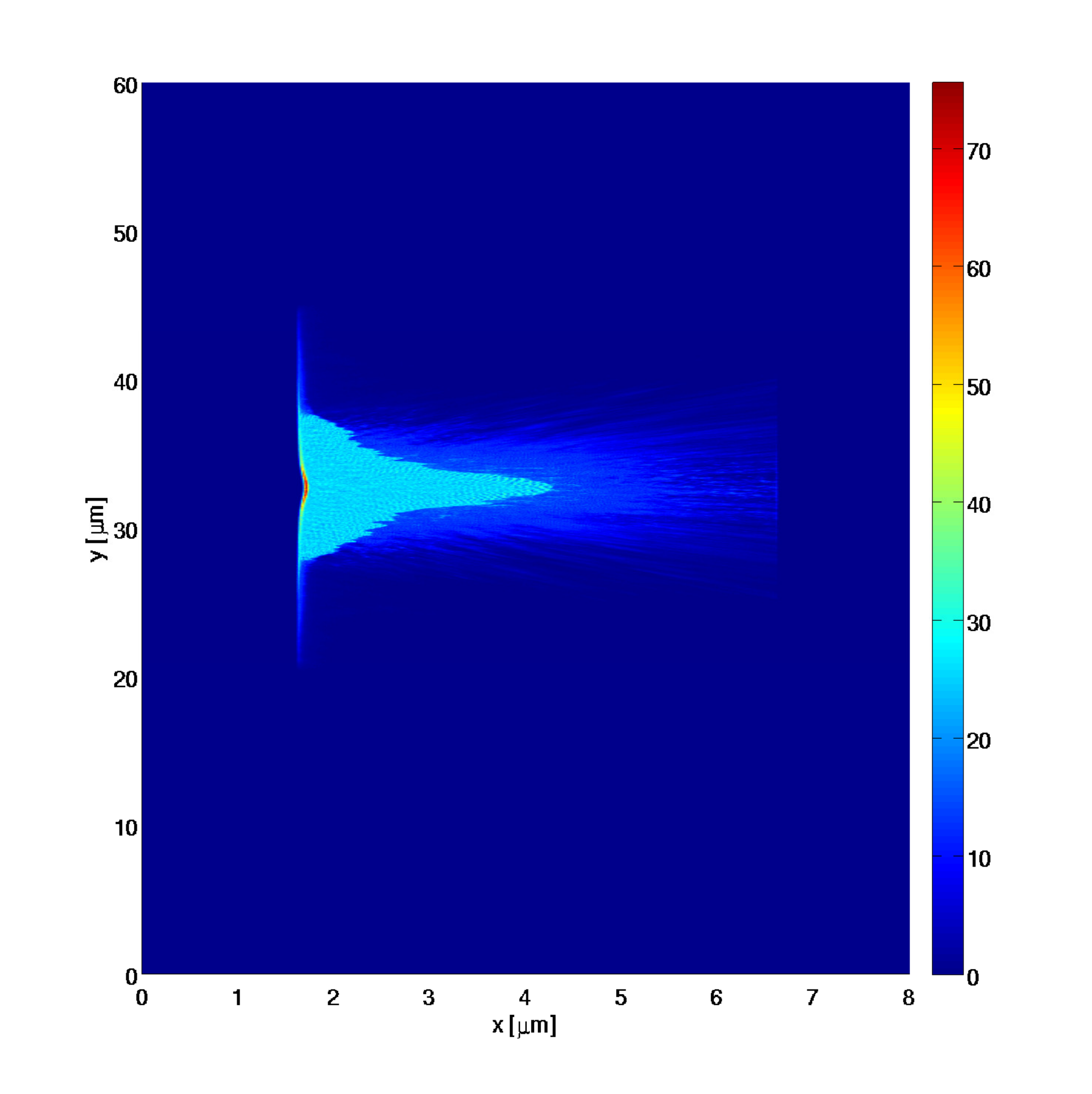}
    
    \subcaption{}
     \end{minipage}
  \hfill
  \begin{minipage}[b]{0.45\textwidth}
    \includegraphics[width=\textwidth]{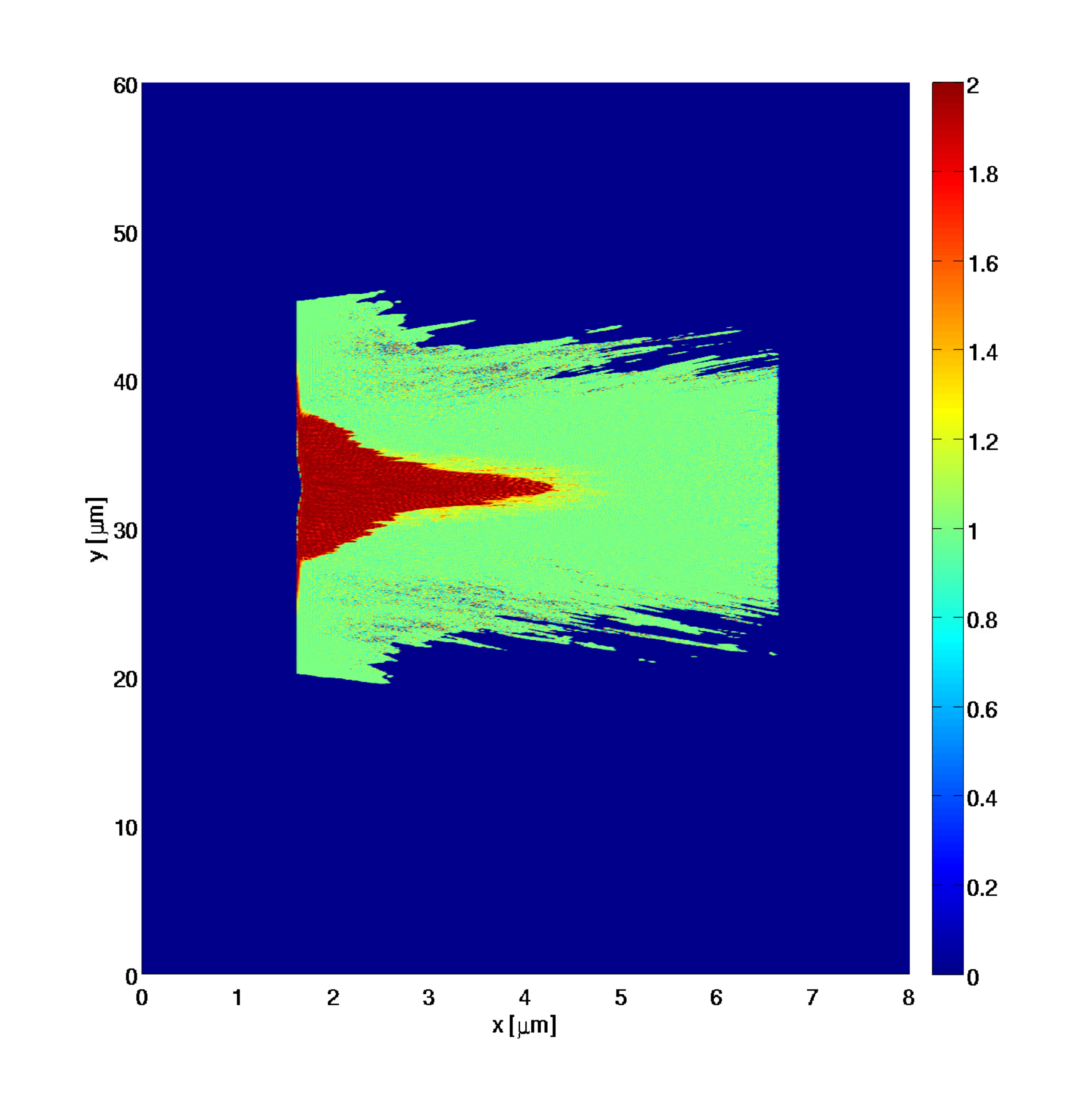}
    \subcaption{}
  \end{minipage}
  \caption{The electron density in units of critical density (a) and the degree of ionization (b) at 100~fs. The laser intensity is $10^{18}$~W/cm$^2$.}
  \label{neZ18b}
\end{figure}

\begin{figure}
\center
\resizebox{0.5\textwidth}{!}{\includegraphics{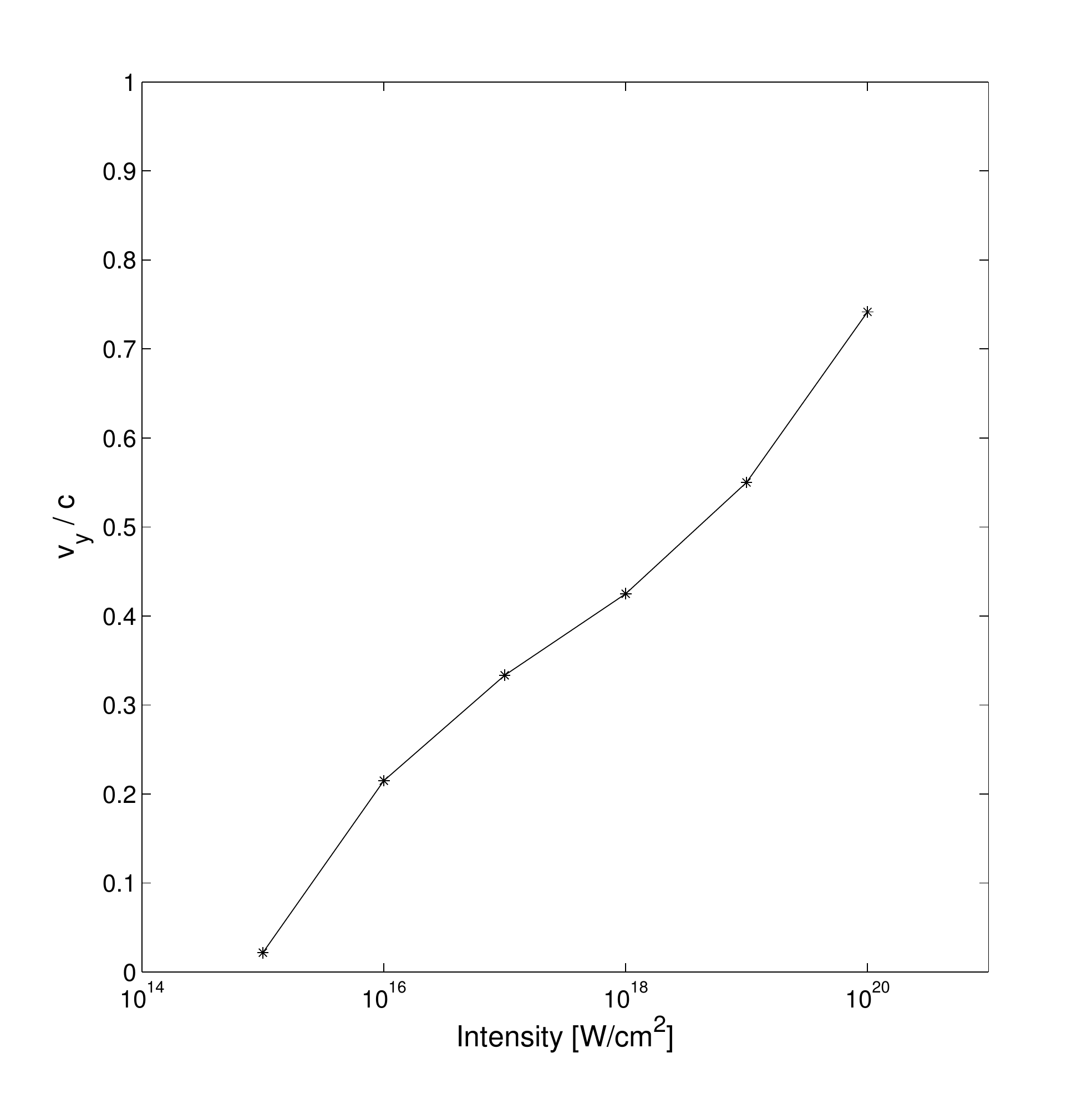}}
\caption{The average speed at 100~fs along the jet divided by the speed of light versus the laser intensity.}
\label{Vyb}
\end{figure}

As a proposal for future experiments on inhomogeneous HDM and to reveal the impact of the jet diameter, we repeated 
the simulation, however, assuming 100 fs pulse length and a $5$~$\mu$m jet. Other parameters are kept the same.
Qualitatively similar results are obtained. For a laser intensity $10^{15}$~W/cm$^2$, a small area within the 
interaction area is weakly ionized. Again, underdense plasmas with a degree of ionization of about one are formed. 
Increasing the laser intensity produces overdense plasmas, see \figurename{~(\ref{neZ18b}~a)}. Consequently, part of 
the laser pulse energy is turned into transient fields. 
Therefore, a new energy absorption window is opened. As clearly seen in \figurename{~(\ref{neZ18b}~b)}, 
the transient field drives ionization along the jet.
The average speed of the ionization-front along the jet as a function of laser intensities 
is displayed in \figurename{~(\ref{Vyb})}.

\subsection{Time evolution of the ionization after the laser pulse}

\begin{figure}
\center
\resizebox{0.5\textwidth}{!}{\includegraphics{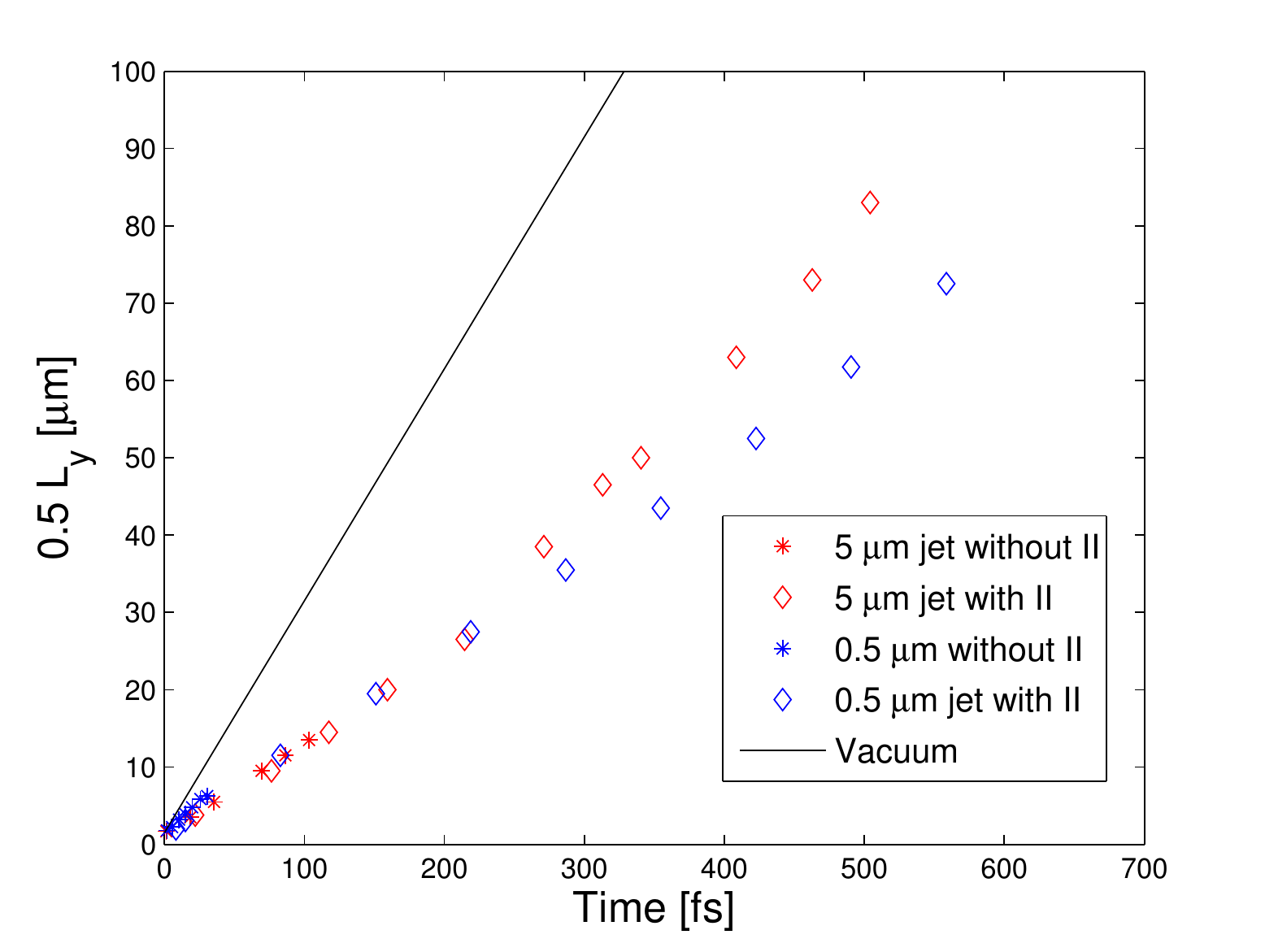}}
\caption{The temporal variation of the ionized area along the jets with and without impact ionization (II), 
the field ionization is considered in all cases. Solid line represents the length corresponding to the speed of light.}
\label{jetab}
\end{figure}

After the laser pulse, the plasma waves (due to the oscillation of critical surfaces and the charge-neutralizing 
disturbance at the jet surface) in the interaction area sustain the transient field,
and the ionization of the jet continues to hundreds of femtoseconds.
\figurename{~(\ref{jetab})} shows the temporal variation of the ionized area length along He jets as well as the 
temporal variation of a length corresponding to the speed of light (solid line).
Here we consider the results of the two examples mentioned in section \ref{A} (represented via blue stars) 
and in section \ref{B} (represented via red stars) with a laser intensity of $10^{18}$~W/cm$^2$.
To reveal the role of impact ionization, we display also in \figurename{~(\ref{jetab})} the results 
of the same two examples (represented via blue and red diamonds), however, considering impact ionization 
in addition to tunnel ionization. 
In order to run the simulations for these two cases until 0.5~ps and due to the limitation of available 
computational resources, we performed the simulations within another configuration: 
$\Delta x= \Delta y = \lambda/50$, $\Delta t=0.5 \Delta x/c$, and 60 super-particles per cell.
Moreover, we increased the width of the simulation domain, 
so that the number of hot electrons, which can escape from the simulation domain, is negligible.
The coincidence between the results considering and neglecting impact ionization proves that this 
process is not important during the laser pulse. 

The main phenomena, we found in our simulations, are the following: (i) In \figurename{~(\ref{jetab})}, one can easily deduce that the ionized area increases with time and the 
ionization-front propagates with speeds in the order of a few tens percent of the speed of light. (ii) The
over-all number of free electrons is still increasing long after the laser pulse. (iii) The degree of ionization is, however, highly inhomogeneous. (iv) Approximately, local equilibrium conditions are fulfilled, however, the electron distribution function of the whole system is strongly non-Maxwellian.

A promising experimental tool to track these time-dependent phenomena could be x-ray Thomson scattering. Therefore, in the next chapter, we will revisit the x-ray Thomson scattering theory
and present some results providing the direct proportionality of the intensity of x-ray Thomson scattering signal and the volume of the ionized region.

%

\newpage

\section{The simulation of a XRTS spectrum}
\subsection{Theoretical basics}

The cross section for a photon with certain wave vector and frequency, ${\bf k}_0, \omega_0$, into a photon state 
with ${\bf k}_1, \omega_1$  is related to the dynamic structure factor (DSF) of all electrons according to:
\begin{equation}\label{cross_section}
\frac{d^2\sigma}{ d \omega d\Omega}=\sigma_{\rm T} S_{ee}^{\rm tot}({\bf k},\omega)
\end{equation}
with the Thomson scattering cross section $\sigma_{\rm T}$
and the dynamical structure factor of the electrons, $S_{ee}^{\rm tot}({\bf k},\omega)$, where ${\bf k}={\bf k}_1-{\bf k}_0$
and $\omega=\omega_1-\omega_0$.

According to Chihara's approach, the DSF can be expressed by a sum of independent contributions
\begin{eqnarray}\label{chihara}
  S_{\rm ee}^{\rm tot}(k,\omega) &=& Z_{\rm f} S_{\rm ee} (k, \omega) + |f(k) +q(k)|^2 S_{\rm ii} (k,\omega) \nonumber \\
  && + Z_{\rm b} \int d\omega' \, S_{\rm c}(k,\omega)\, S_{\rm s}(k,\omega-\omega'). 
\end{eqnarray}
The first term denotes the DSF of free electrons with $Z_{\rm f}$ being
the number of free electrons per nucleus. The second term characterizes the scattering from weakly and tightly bound electrons.
Its amplitude is determined by the sum of the form factor $f (k)$ of bound electrons and the screening cloud $q(k)$ of free electrons; 
the ion-ion structure factor, $S_{\rm ii} (k,\omega)$, which represents the thermal motion of the ions.
The last term is due to the contribution of the bound-free transitions $S_{\rm c}$.
This term describes Raman-type transitions of inner shell electrons to the continuum which are modulated by the ion motion contained in 
$S_{\rm s}(k,\omega)$ and multiplied by the core charge $Z_{\rm b}$.

Here we will concentrate on the free electrons' contribution. In thermodynamic equilibrium, the DSF of free electrons 
can be expressed via the fluctuation-dissipation theorem by the inverse dielectric function $\epsilon^{-1}(k,\omega)$:
\begin{equation}\label{See0}
  S_{\rm ee}=-\frac{\epsilon_0 \hbar k^2}{\pi e^2 n_{\rm e}} \frac{{\rm Im} \,\, \epsilon^{-1}(k,\omega)}{1-\exp(\frac{-\hbar \omega}{k_{\rm B} T_{\rm e}})}\,.
\end{equation}
This relation is used also for nonhomogeneous situations with local thermodynamic equilibrium conditions. 
Then the Thomson scattering signals of different volume elements with local temperature and densities are summed up. 
The dielectric function is usually calculated in random phase approximation or generalizations like the Born-Mermin 
approximation~\cite{Theile2010}.

In non-equilibrium situations with stronger deviations of the electron distribution from an equilibrium form, 
Eq.~(\ref{See0}) is not applicable. Instead, one can use the non-equilibrium random phase approximation 
expression~\cite{Kraeft,Chapman2011,Chapman2012,Cordes2016} for the DSF, 
$S_{ee}(\textbf{k},\omega,t) = S_{ee}^{0}(\textbf{k},\omega,t)/|\varepsilon^{RPA}(\textbf{k},\omega,t)|^{2}$, with 
\begin{eqnarray}
 S_{ee}^{0}(\textbf{k},\omega,t)&=&\frac{2\hbar}{n_{e}}\int \frac{\mathrm{d}^{3}\textbf{p}}{(2\pi\hbar)^{3}} \delta(E(\textbf{p}+\textbf{k})-E(\textbf{p})-\hbar \omega)\nonumber \\
 &&\times [1-f_{e}(\textbf{p}+\textbf{k},t)]f_{e}(\textbf{p,t})
\end{eqnarray}
and the dielectric function $\varepsilon^{RPA}(\textbf{k},\omega,t)=1-V_{\rm ee}(k)\chi^{0}_{\rm ee} (\textbf{k},\omega,t)$.
We have $V_{\rm ee}(k)=4\pi\hbar^{2}e^{2}/k^{2}$, and $\chi^{0}_{\rm ee}$ is the response function of the non-interacting system,
\begin{equation}
 \chi^{0}_{\rm ee}(\textbf{k},\omega,t)=2\int \frac{\mathrm{d}^{3}\textbf{p}}{(2\pi \hbar)^{3}} \frac{f_{e}(\textbf{p}+\textbf{k},t)-f_{e}(\textbf{p},t)}{E(\textbf{p}+\textbf{k})-E(\textbf{p})-\hbar \omega - i0}\,.
 \label{polfunction}
\end{equation}
All expressions are valid for non-equilibrium distributions functions $f_{\rm e}(\textbf{p},t)$,
being normalized as follows
$n_{\rm e}=2 \int\frac{{\rm d}^3 p}{(2\pi\hbar)^3} f_e({\bf p})\,.$

For nondegenerate particles, the Pauli blocking term $(1-f)$ can be neglected:
\begin{eqnarray}
S^0_{\rm ee}(k,\omega)=\frac{2\hbar}{n_{\rm e}}\int\frac{{\rm d}^3 p}{(2\pi\hbar)^3}
f_{\rm e}({\bf p}) \, \delta\left(\hbar \omega+\frac{{\bf p}\cdot \hbar{\bf k}}{m_{\rm e}}+\frac{\hbar^2 k^2}{2m_{\rm e}}\right)
\,.
\end{eqnarray}
The delta function describes Doppler broadening and Compton shift. It can be used to perform one angle integral with $z=\cos{\theta}$. Because $|z|<1$, this implies $p>|\frac{m_a \omega}{k}+\frac{\hbar k}{2}|$.
\begin{eqnarray}\label{noncollective}
S^0_{\rm ee}(k,\omega)=\frac{2\hbar}{n_{\rm e}}\frac{2\pi m_{\rm e}}{\hbar k\,(2\pi\hbar)^3} \int\limits^\infty_{|\frac{m_{\rm e} \omega}{k}+\frac{\hbar k}{2}|} {\rm d}p\, p\, f_{\rm e}({p}) \,.
\end{eqnarray}
This equation is especially useful for the non-collective scattering regime. Due to $|\varepsilon(k,\omega)|\approx 1$, the scattering signal is determined by $S^0_{\rm ee}(k,\omega)$ only and,
therefore by the electron distribution function of the whole scattering volume. No assumptions on local temperatures are necessary.
The quantum diffraction term, $\hbar k/(2m_{\rm e})$, connected with the Compton shift, $-\omega=\hbar k^2/(2m_{\rm e})$, is responsible for an asymmetry of the scattering signal.

For a Maxwell-Boltzmann distribution function, i.e.\ in equilibrium, the integral can be performed analytically to give
\begin{eqnarray}
S^{0}_{\rm ee}(k,\omega)&=&\frac{m_{\rm e}}{\hbar k\sqrt{2\pi m_{\rm e} k_B T}}\exp{\left[-\frac{m_{\rm e}\left(\frac{\omega}{k}+\frac{\hbar k}{2m_{\rm e}}\right)^2}{2 k_BT}\right]}
\,.
\end{eqnarray}
The asymmetry of this (equilibrium) function is connected with the detailed balance condition
\begin{eqnarray}
\frac{S^{0}_{\rm ee}(k,-\omega)}{S^{0}_{\rm ee}(k,\omega)}=\exp{\left(\frac{\hbar \omega}{k_B T}\right)}
\,.
\end{eqnarray}
Because $|\varepsilon(k,-\omega)|=|\varepsilon(k,\omega)|$, the detailed balance equation is also valid
for $S_{\rm ee}(k,\omega)$.

\subsection{Calculation of synthetic XRTS spectra for the simulation case}

\begin{figure}[ht]
\center
\resizebox{0.5\textwidth}{!}{\includegraphics{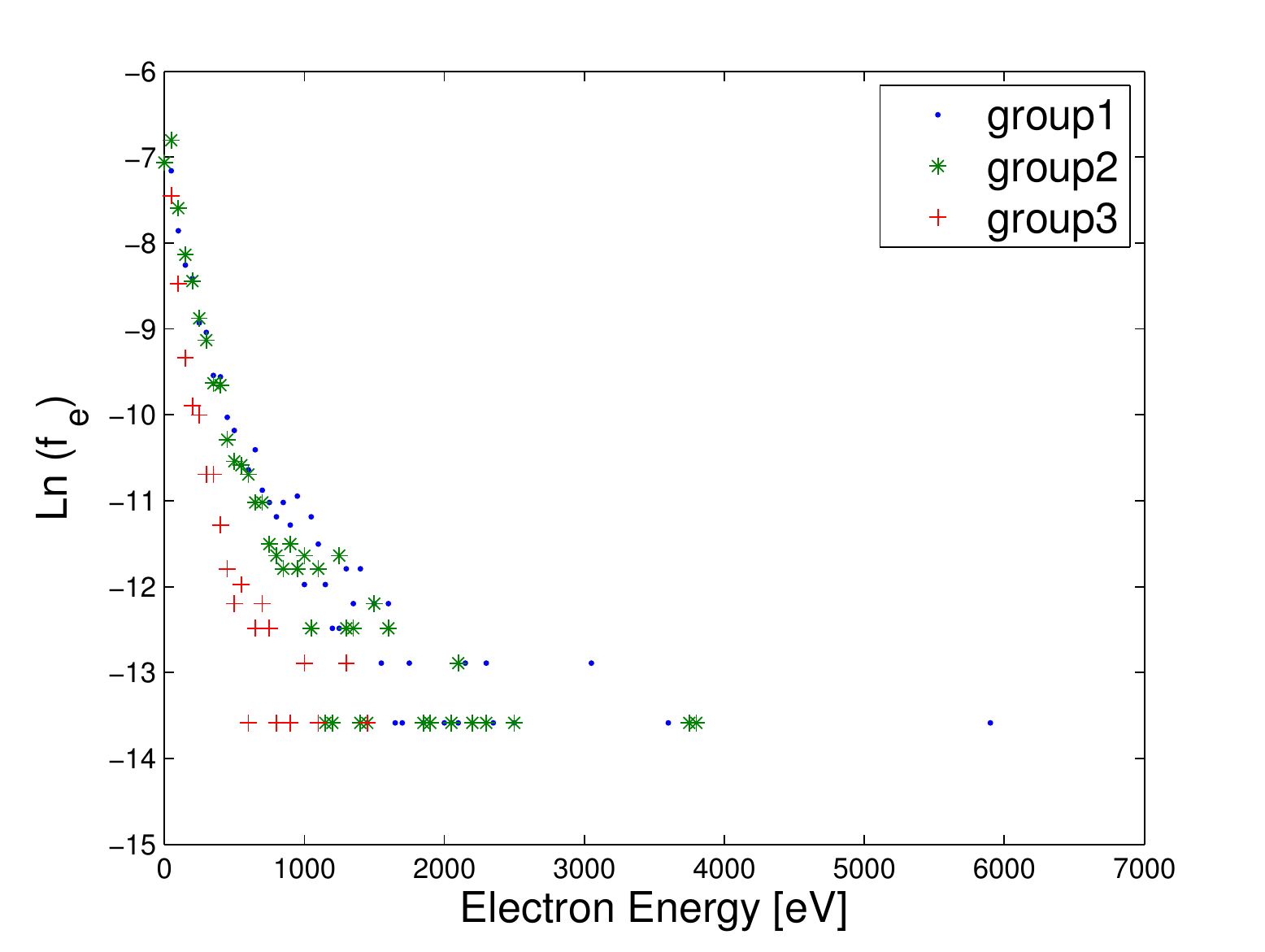}}
\caption{The electron energy distribution in some group of cells.}
\label{lnfecell}
\end{figure}

\begin{figure}[!tbp]
  \centering
  \begin{minipage}[b]{0.45\textwidth}
    \includegraphics[width=\textwidth]{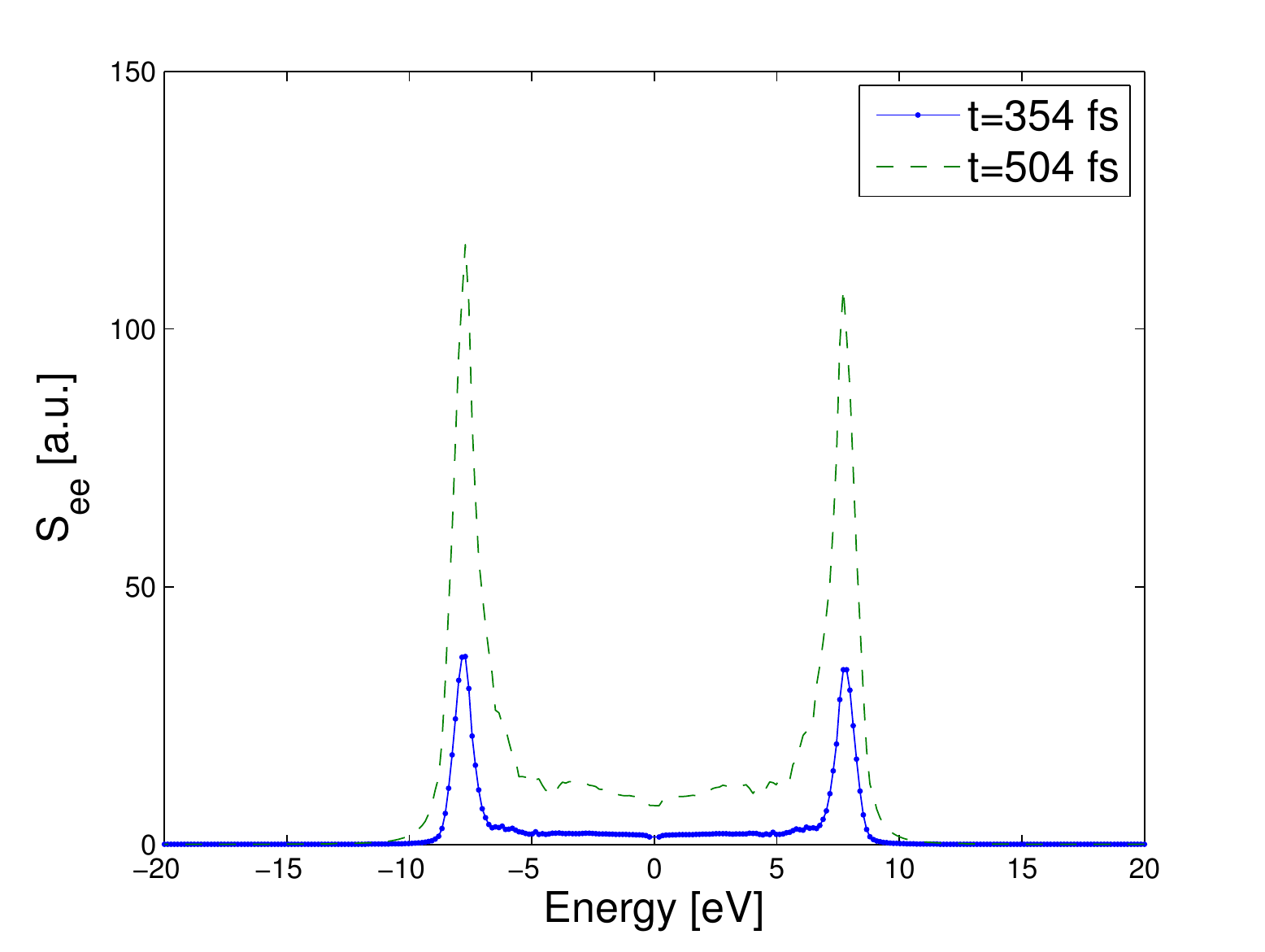}
    \subcaption{}
      \end{minipage}
  \hfill
  \begin{minipage}[b]{0.45\textwidth}
    \includegraphics[width=\textwidth]{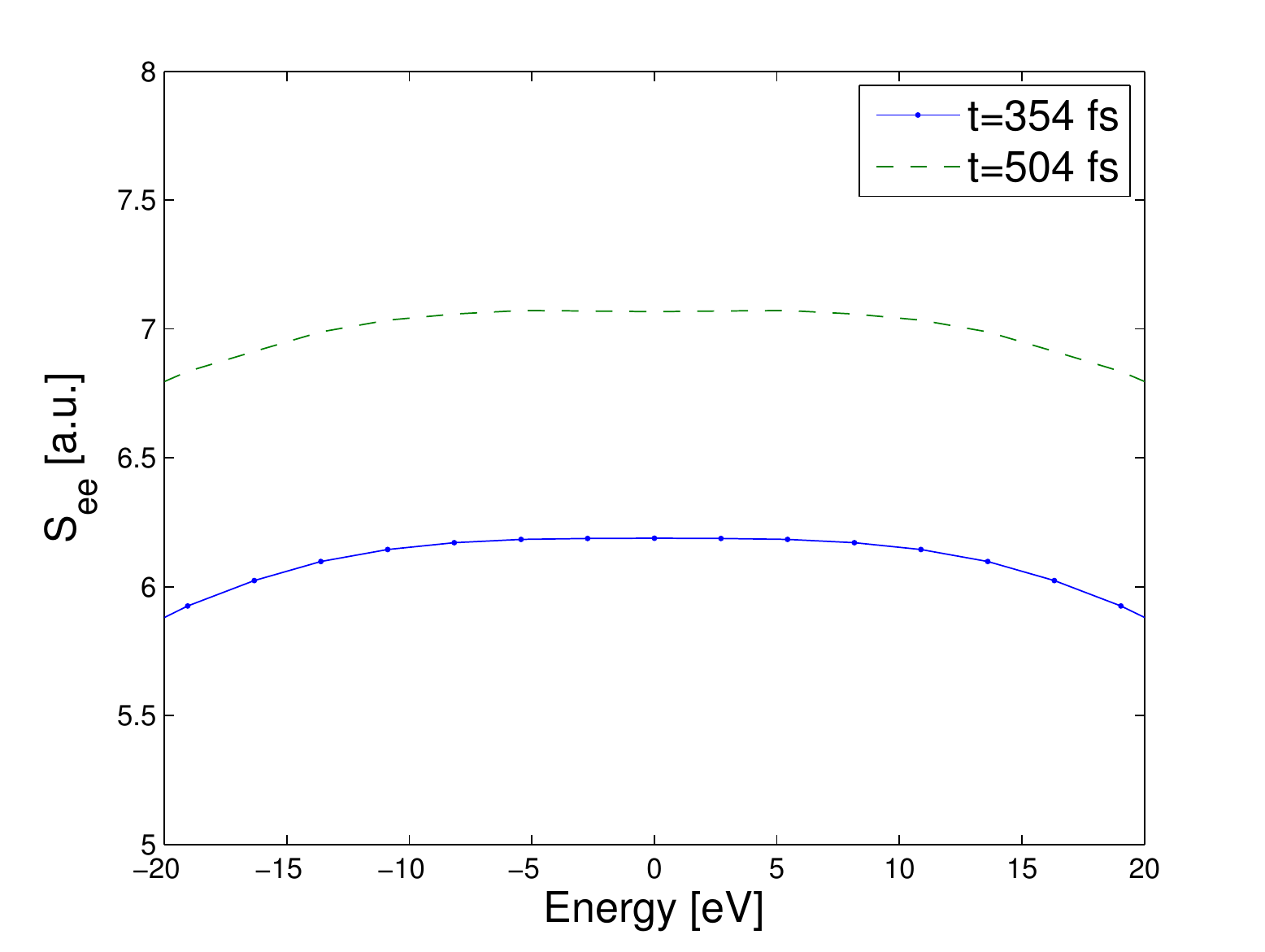}
    \subcaption{}
  \end{minipage}
  \caption{Collective (a) and non-collective (b) Thomson scattering spectrum of free electrons of the 5~$\mu$m jet considering impact 
ionization beside field ionization, a laser intensity of $10^{18}$~W/cm$^2$, and a pulse length of 100~fs 
at 354~fs and 504~fs.}
  \label{See}
\end{figure}

The PIC simulation results of the cryogenic He jet pumped by the intense ultrashort laser pulse show highly
inhomogeneous conditions for the ionization degree and free electron density. Concerning the use of local 
temperatures one has to be aware of the limitations: PIC simulations require fine spatiotemporal grids; 
moreover, the number of super-particles in each cell is finite. Therefore, the electron distribution cannot 
be fitted in each cell with Maxwell-Boltzmann distributions. Instead, we look for larger, semi-homogeneous 
regions (i.e.\ group of cells in the PIC simulation) and approximate the electron distribution by a 
Maxwell-Boltzmann distribution with an effective temperature  $k_{\rm B} T_{\rm eff}=2/3 {\bar E}$. 
The distributions functions in some regions have a noisy, but very hot tail. Thus, to avoid the 
overestimation of the electron temperature, we excluded in the calculation of the mean energy all 
electrons with energies more than 1 keV. However, we found that the cut-off at 1000~eV is appropriate 
from the statistical point of view for most of the cells, as shown in \figurename{~(\ref{lnfecell})}. 
We would like to emphasis that changing the arbitrary cut-off will change the estimated temperature, 
but the results in \figurename{~(\ref{See})} will be qualitatively the same. Especially, hot electrons 
will not contribute to collective plasma excitations, the plasmons.  


Therefore, for inhomogeneous systems with local thermodynamic equilibrium conditions, the scattering 
signal can be theoretically evaluated by a sum of the different cells. This applies especially for the 
collective scattering regime, i.e.\ a probe laser wavelength $\lambda_0=13.5$~nm and for a scattering 
angle of $60^{\circ}$. We use the plasma parameter profiles from the PIC simulations of the 5~$\mu$m 
jet with impact ionization at 345~fs and 504~fs (profiles related to results represented by the red 
diamonds in \figurename{~(\ref{jetab})}). To achieve better statistics, the temperature of electrons 
and ions in each cell ($8\Delta x \times 128\Delta y$) is determined according to the Maxwell-Boltzmann 
distribution via the particle energies. 

The overall Thomson scattering due to free electrons is displayed in \figurename{~(\ref{See}~a)}. 
As shown, the amplitude of the plasmons increases by time due to the propagation of the transient 
field and the increase in the ionized volume. We want to note that in inhomogeneous systems, despite 
that each cell or group of cells obeys approximately local thermodynamic equilibrium, the total system 
might be far away from equilibrium. Hence, the system in total has not a Maxwell-Boltzmann distribution 
and attributing one effective temperature to the system is questionable.


In the non-collective scattering regime, no assumption on local thermodynamic equilibrium is necessary 
provided the velocity distribution function $f_{\rm e}(p)$ of the whole scattering volume is known from 
the PIC simulations. In this regime we have $\arrowvert \epsilon^{R} (k,\omega)\arrowvert ^2 \approx 1$ 
and $S_{\rm ee} \approx S^0_{\rm ee}$ determined by Eq.~(\ref{noncollective}).
In order to calculate the non-collective scattering off the jet, we consider a probe pulse with a 
wavelength of $2$~nm and a scattering angle of $160^{\circ}$. 
\figurename{~(\ref{See}~b)} displays the non-collective scattering of free electrons at the 
same delay times as in \figurename{~(\ref{See}~a)}. Again, as it has been concluded previously, the 
intensity of the spectrum at 504~fs is larger than that at 354~fs due to the increase in the ionized volume.
Moreover, the qualitative behavior of the results in \figurename{~(\ref{See}~b)} depends only 
weakly on $k$ provided the criteria of non-collective scattering are fulfilled.

\section{Conclusion}

We studied the interaction of ultrashort and intense laser pulses with cryogenic He jets. 
For intensities $\ge 10^{16}$~W/cm$^2$, the laser pulse has been found to produce overdense 
plasmas and strong transient field sufficient to ionize the jet outside the interaction area. 
The ionization front propagates along the jet with a fraction of the speed of light. 
An appropriate experimental method to infer this ionization dynamics in the jet after 
excitation by an intense laser pulse could be x-ray Thomson scattering. To illustrate this, 
synthetic XRTS spectra have been calculated for different delay times after the laser pulse 
and different scattering conditions. The amplitude of the plasmons in the collective scattering 
regime has been found to increase as a function of delay time whereas the position does not change much. 
The intensity of the non-collective scattering signal increases as well with the time, 
connected with the increase of the ionized volume and, therefore, the number of scatterers. 
Beside the variation of the time delay between pump and probe pulse, it might useful to vary 
the focus of the probe beam in order to get information on the spatial extension of the ionized area.

We want to remark that the transient fields inside the plasma which cause ionization outside the 
interaction area is a result of kinetic instabilities as two-plasmon decay, oscillation of critical surfaces, 
and the charge disturbance at the jet surface due to the escape of hot electrons. 
All these phenomena might be interrupted with hydrodynamic instabilities due to the ion motion 
on ps time scales. The kinetic simulation up to times on the ps scale is demanding, 
and further 2d or 3d hydrodynamics simulation is needed.

\section{Acknowledgement}
We thank P. Sperling, U. Zastrau, C. Roedel, and W-D. Kraeft for interesting discussions.
This study was supported by the DFG within the SFB~652 and by the BMBF within the FSP~302. \\


\end{document}